\documentclass[12pt]{article}

\pdfoutput=1
\usepackage{slashed}
\usepackage{color, verbatim}
\usepackage{latexsym}
\usepackage{epsfig}
\usepackage{slashed}
\usepackage{amsmath,accents}

\usepackage{amssymb}
\usepackage{graphicx}
\usepackage{bm}

\usepackage[font={small}]{caption}

\usepackage{cite}
\usepackage{hyperref}

\definecolor{myred}{rgb}{0.7, 0, 0}
\definecolor{myblue}{rgb}{0, 0, 0.7}
\definecolor{mygreen}{rgb}{0.04, 0.7, 0.5}
\hypersetup{colorlinks,citecolor=myred,linkcolor=myblue,urlcolor=myblue,linktocpage=true}

\makeatletter

\@addtoreset{equation}{section}
\makeatother

\def\lsim{\mathrel{\raise.3ex\hbox{$<$\kern-.75em\lower1ex\hbox{$\sim$}}}}
\def\gsim{\mathrel{\raise.3ex\hbox{$>$\kern-.75em\lower1ex\hbox{$\sim$}}}}

\newcommand{\be}{\begin{equation}}
\newcommand{\ee}{\end{equation}}
\newcommand{\bea}{\begin{eqnarray}}
\newcommand{\eea}{\end{eqnarray}}

\begin{document}

\thispagestyle{empty}

\begin{center}

\begin{center}

{\LARGE\sf Higgsino Dark Matter in the MSSM \\ with gravity mediated SUSY breaking}

\end{center}

\vspace{.5cm}

\textbf{
 Antonio Delgado$^{\,a}$, Mariano Quir\'os$^{\,b}$
}\\

\vspace{1.cm}

${}^a\!\!$ {\em {Department of Physics, University of Notre Dame, 225 Nieuwland Hall \\ Notre Dame, IN 46556, USA
}}

${}^b\!\!$ {\em {Institut de F\'{\i}sica d'Altes Energies (IFAE) and BIST, Campus UAB\\ 08193, Bellaterra, Barcelona, Spain
}}

\end{center}

\begin{quote}\small
\begin{center}\textsc{Abstract} \end{center}

A cosmologically stable neutral component from a nearly pure $SU(2)$ doublet, with a mass $\sim$1.1 TeV, is one appealing candidate for dark matter (DM) consistent with all direct dark matter searches.  We have explored this possibility in the context of the Minimal Supersymmetric extension of the Standard Model (MSSM), with the Higgsino playing the role of DM, in theories where supersymmetry breaking is transmitted by gravitational interactions at the unification scale $M\simeq 2\times 10^{16}$ GeV. We have focussed our work in the search of ``light" supersymmetric spectra, which could be at reach of present and/or future colliders, in models with universal and non-universal Higgs and gaugino Majorana masses. The lightest supersymmetric particles of the spectrum are, by construction, two neutralinos and one chargino, almost degenerate, with a mass $\sim $1.1 TeV, and a mass splitting of a few GeV. Depending on the particular scenario the gluino can be at its experimental mass lower bound $\sim$ 2.2 TeV; in the squark sector, the lightest stop can be as light as $\sim$ 1.6 TeV, and the lightest slepton, the right-handed stau, can have a mass as light as $1.2$ TeV. The lightest neutralino can be found at the next generation of direct dark matter experimental searches. In the most favorable situation, the gluino, with some specific decay channels, could be found at the next run of the Large Hadron Collider (LHC), and the lightest stop at the High-Luminosity LHC run.

\end{quote}
\vfill
  
\newpage
\section{Introduction}

Supersymmetry, and in particular the Minimal Supersymmetric extension of the Standard Model (MSSM), remains as the most appealing solution to the Standard Model grand naturalness problem~\cite{Haber:1984rc,Martin:1997ns}. In spite of all the negative results from experimental searches, the fact that the Higgs boson was found with a mass $m_h\simeq 125$ GeV, points toward a heavy supersymmetric spectrum, so that nature should be affected by an irreducible little hierarchy problem to live with. Supersymmetric spectra in the few TeV range are still allowed by present searches at the Large Hadron Collider (LHC)~\cite{PDG}.

An important spin-off of the MSSM, in the presence of $R$-parity conservation, which prevents baryon and lepton number violation at the perturbative level and, thus, proton decay, is that the Lightest Supersymmetric Particle (LSP) is a candidate for (cold) dark matter (DM), if it is electrically neutral, and can give rise to the observed cosmological abundance of dark matter density after thermal freeze-out~\cite{Jungman:1995df}. The possibility that DM is described as the lightest neutralino has been explored since long ago in the literature as one of the most appealing features of the MSSM~\cite{ArkaniHamed:2006mb,Baer:2011ab,Low:2014cba,Roszkowski:2014wqa,Roszkowski:2014iqa,Roszkowski:2017nbc,Cheung:2012qy,Martin:2007gf,Hall:2012zp,Giudice:2010wb,Boehm:2013qva,Fowlie:2013oua,Huang:2017kdh,Badziak:2017the}.
Given the strong bounds on the mass of supersymmetric particles, and the plethora of null results from direct search experiments~\cite{PDG}, there is a clearly preferred scenario: a nearly pure Higgsino with a mass $\sim 1.1$ TeV~\cite{Roszkowski:2017nbc,Kowalska:2018toh}.

In view of the previous comments, we will consider in this paper the possibility for the MSSM to encompass a supersymmetric spectrum where the LSP is a nearly pure Higgsino with a mass $\sim 1.1$ TeV. As the MSSM spectrum does largely depend on the supersymmetry breaking mechanism, and on the solution to the supersymmetric $\mu$-problem (generation of the $\mu$ term in the Higgs superpotential), we will do it in models of gravity mediation of supersymmetry breaking, where the $\mu$ term can be generated through non-renormalizable contribution to the K\"ahler potential, the Giudice-Masiero  mechanism~\cite{Giudice:1988yz}. For these models the scale $M$ at which supersymmetry is broken (i.e.~the scale at which the soft breaking masses are generated) is identified with the scale where gauge couplings unify, i.e.~the unification scale $M\simeq 2\times 10^{16}$ GeV~\cite{Martin:1997ns}. These models have minimal supergravity as the ultraviolet (UV) completion and are inspired by, and obtained from, string constructions~\cite{Brignole:1998dxa}. In these scenarios the soft breaking masses do depend on the localization of Standard Model fields in the extra dimensions, so that two simple scenarios for scalars are: i) Models where, at the unification scale, Higgs and sfermion masses are equal (dubbed universal Higgs mass models) and, ii) Models where Higgs and sfermion masses are different (dubbed non universal Higgs mass models). On the other hand, as gaugino Majorana masses ($M_a$) evolve as the corresponding gauge couplings ($\alpha_a$), it is usually assumed that gaugino Majorana masses unify at the gauge coupling unification scale, although, in supergravity models these masses depend on non-renormalizable $F$-density couplings in the gauge sector and can be different. 

To summarize, the main purpose of this paper is to make predictions on the supersymmetric mass spectra of models where a 1.1 TeV Higgsino is the LSP, and supersymmetry is broken at the unification scale, which can be useful to guide experimental searches and, in particular, to seek the existence of supersymmetric spectra which will be at reach of present or future colliders. Given the present bounds on supersymmetric masses, we will not pay particular attention to the issue of fine-tuning~\footnote{An analysis based on fine-tuning criteria was done in Ref.~\cite{Kowalska:2014hza}.} but, instead, on the possibility of experimental detection of the supersymmetric spectra. For that reason we will gave up the criterium of Majorana mass unification and consider cases where the gluino is on the verge of experimental detection. We will see that, with a lighter gluino  the renormalization over the other supersymmetric parameters is smaller and the resulting squark spectra are lighter than those with a heavy gluino.

The contents of this paper are as follows. In Sec.~\ref{sec:breaking} the conditions for electroweak and supersymmetry breaking are summarized for the scalar sector. In Sec.~\ref{sec:dm} the conditions for the LSP to be an almost pure Higgsino with a mass of $\sim 1.1$ TeV are established. The spectrum of charginos and neutralinos is fully determined, with all generality, after the conditions from the XENON1T direct searches are imposed. In Sec.~\ref{sec:susybreaking} the predictions on supersymmetric spectra for different scenarios of supersymmetry breaking are obtained. In particular, scenarios of universal and non-universal Higgs masses, at the unification scale $M$, as well as those of universal and non-universal gaugino masses, will be separately studied. In Sec.~\ref{sec:phenomenology} some comments on the experimental signatures at hadron colliders of the considered scenarios are pointed out. Finally in Sec.~\ref{sec:conclusion} the conclusions and outlook are drawn.

\section{Electroweak and Supersymmetry Breaking}
\label{sec:breaking}
In the MSSM electroweak symmetry breaking (EWSB) is achieved by means of two Higgs doublets $H_U$ and $H_D$ whose vacuum expectation values (VEVs) give a mass to up-like quarks, and down-like quarks and charged leptons, respectively. The corresponding superfields $\mathcal H_U$ and $\mathcal H_D$ appear in the superpotential as
$
W=\mu \mathcal H_U\cdot \mathcal H_D\,,
$
which gives a supersymmetric mass to Higgs bosons and Higgsinos. Moreover, through the process of supersymmetry breaking, the Higgs sector acquires soft-breaking masses as
\be
-\mathcal L_{\rm soft}=m_{H_U}^2|H_U|^2+m_{H_D}^2|H_D|^2+\left(b\, H_U\cdot H_D+h.c.\right)
\ee

After imposing EW breaking at a low scale $\mathcal Q_0$ as $\langle H_U\rangle=v_U$, $\langle H_D\rangle =v_D$, with $t_\beta\equiv\tan\beta=v_U/v_D$, the equations of minimum (EoM) are found as
\begin{align}
m_{H_U}^2&=-\left(\mu^2+\frac{1}{2}m_Z^2\right)\frac{t_\beta^2-1}{t_\beta^2}+m_{H_D}^2/t_\beta^2\nonumber\\
\sin 2\beta&=\frac{2b}{m_{H_U}^2+m_{H_D}^2+2\mu^2}\label{eq:EoM}
\end{align}
where all parameters are considered at the scale $\mathcal Q_0$.

As it is obvious from the EoM~(\ref{eq:EoM}), EWSB in the MSSM requires supersymmetry breaking. We will assume that at the high scale $M$, the supersymmetry breaking scale, soft breaking parameters are generated for the Higgses, gauginos and (third generation) squarks as
\be
m_{H_U}^0,\quad m_{H_D}^0,\quad m_Q^0,\quad m_U^0,\quad m_D^0,\quad A_t^0,\quad M_a^0\quad (a=1,2,3)
\label{eq:BC}
\ee
where the zero upper index indicates that the corresponding parameter is evaluated at the scale $M$. Using the renormalization group evolution of the parameters from the high scale $M$ to the low scale $\mathcal Q_0$, the values specified in Eq.~(\ref{eq:BC}) should be considered as boundary conditions. We will specify the corresponding parameter values at the low scale $\mathcal Q_0$ with no upper index,  i.e.~$m_{X}$.

We will, hereafter, consider gravity-mediated supersymmetry breaking, for which the supersymmetry breaking scale is at the unification scale $M\sim 2\times 10^{16}$ GeV, and the soft breaking terms do depend on the superpotential and K\"ahler potential dependences of the superfield $\mathcal X$ which spontaneously break supersymmetry through its $F_X$-term. This allows many different possibilities, or relationships, between the supersymmetry breaking parameters in (\ref{eq:BC})~\cite{Martin:1997ns}. Moreover in supergravity models the $\mu$ and $b$ terms can be obtained via the Giudice-Masiero mechanism~\cite{Giudice:1988yz}, through non-renormalizable contributions to the K\"ahler potential as
\be
K=\frac{\lambda_\mu}{M_P} \mathcal H_U\cdot \mathcal H_D \mathcal X^\dagger+\frac{\lambda_b}{M_P^2}\mathcal H_U\cdot \mathcal H_D |\mathcal X|^2+h.c.
\ee
leading to
\be
\mu^0=\frac{\lambda_\mu}{M_P}F_X^\dagger,\quad b^0=\frac{\lambda_b}{M_P^2}|F_X|^2
\label{eq:GM}
\ee
As the values of the $\mu^0$ and $b^0$ terms at the scale $M$ do depend on unknown parameters of the UV supergravity completion, we can consider their values at the low scale $\mathcal Q_0$ in the EoM as free parameters (\ref{eq:EoM}): $\mu$ and $b$.

Moreover, the gaugino Majorana mass entry $\mathcal M_{ab}$ is given in terms of the K\"ahler potential $K$, superpotential $W$, and the gauge kinetic function $f_{ab}(\phi^i)$, an analytic function of the scalar fields $\phi^i$ which transforms under the gauge group as the symmetric product of adjoint representations, as
\be
\mathcal M_{ab}=\frac{1}{2 Re f_{ab}}e^{-G/2}G^i(G^{-1})_{ij} \frac{\partial f_{ab}^\ast}{\partial\phi_j}
\ee
where $G=K+W$. Depending on the particular UV (supergravity) completion of the model, the gaugino mass spectrum can behave in different ways at the unification scale. A survey of non-universal gaugino mass models from grand unified and string models can be found in Ref.~\cite{Horton:2009ed}.

Here we will mainly study the case of \textit{intermediate} $\tan\beta$, $1\ll \tan\beta\ll m_t/m_b$, so that we will neglect all Yukawa couplings, except the top-quark one. Then we can write the soft breaking terms which appear in the first equation of (\ref{eq:EoM}) in terms of their values at $M$. In Refs.~\cite{Delgado:2014vha,Delgado:2014kqa} we  integrated the renormalization group equations (RGE) between the high scale $M$ and the low scale $\mathcal Q_0$. For cases where the hypercharge $D$-term vanishes, i.e.~$m_{H_U}^2-m_{H_D}^2+\sum_a(m_Q^2-2m_U^2+m_D^2-m_L^2+m_E^2)_a=0$, where $a$ is a generation index, an equality which is RGE invariant (and which will cover all cases considered in this paper), the soft breaking terms, sensitive to the top Yukawa coupling, at the scale $\mathcal Q_0$ are linear combinations of the parameters at the scale $M$: $(m_Q^0)^2$, $(m_U^0)^2$, $(m_{H_U}^{0})^2$, $M_a^0 M_b^0$, $M_a^0 A_t^0$ and $(A_t^{0})^2$. In particular we can write
\begin{align}
m_{H_U}^2&=(m_{H_U}^0)^2+\eta_Q\left[(m_{Q}^0)^2+(m_{U}^0)^2+(m_{H_U}^0)^2\right]\nonumber\\
&+\eta_a\sum_a(M_a^0)^2+\eta_{ab}\sum_{a\neq b}M_a^0M_b^0+\sum_a \eta_{aA}M_a^0 A_t^0+\eta_A (A_t^0)^2
\label{eq:mHU}
\end{align}
where all the coefficients $\eta_X=\eta_X(\mathcal Q_0,M)$ are functions of the high scale $M$ and the low scale $\mathcal Q_0$, fitted 
 in Ref.~\cite{Delgado:2014vha}, and used throughout this work.  As for the other breaking parameter in Eq.~(\ref{eq:EoM}), $m_{H_D}^2$, as we are neglecting the bottom Yukawa coupling, it is renormalized by gauge interactions, so that in the one-loop approximation it can be given by~\cite{Martin:1997ns}
\be
m_{H_D}^2=(m_{H_D}^0)^2+\frac{3}{2}\left(1-\frac{\alpha_2^2(\mathcal Q_0)}{\alpha_2^2(M)}  \right)(M_2^0)^2+\frac{1}{22}\left(1-\frac{\alpha_1^2(\mathcal Q_0)}{\alpha_1^2(M)}  \right)(M_1^0)^2
\label{eq:mHD}
\ee

Concerning the second equation in (\ref{eq:EoM}), the value of $t_\beta$ is determined by the soft-breaking parameter $b$. In the limit of large $t_\beta$, and using the first equation in (\ref{eq:EoM}) it is given by
\be
t_\beta\simeq \frac{m_{H_D}^2+\mu^2-m_Z^2/2}{b}\simeq \frac{m_{H_D}^2}{b}
\ee
where, in the last equality, we are assuming soft breaking masses to be much larger than $\mu$. So, as it is natural in the mechanism of Eq.~(\ref{eq:GM}), for $\lambda_b=\mathcal O(\lambda_\mu)$ we should get $b\simeq \mu^2$ and for $m_{H_D}\gg |\mu|$ we should get $t_\beta\gg 1$. In view of our ignorance on the UV completion of the model we will consider $t_\beta$ as a free parameter.

\section{The Dark Matter sector}
\label{sec:dm}
The Higgs bosons supersymmetric partners, $\tilde H_U$ and $\tilde H_D$, along with the supersymmetric partners of the $SU(3)\otimes SU(2)_L\otimes U(1)_Y$ gauge bosons, $\tilde g$, $\tilde W$ and $\tilde B$, make a set of four neutral Majorana fermions (neutralinos) $\chi^0_{1,2,3,4}$, two charged fermions (charginos) $\chi^\pm_{1,2}$, and eight gluinos $\tilde g$.

Charginos and neutralinos get masses from the superpotential $W=\mu \mathcal H_U\cdot \mathcal H_D$, from the soft-breaking Majorana masses $M_{1,2}$ for $\tilde B$ and $\tilde W$, respectively, and from the electroweak breaking.
 The mass matrices for neutralinos and charginos are then given by
\begin{equation}
\mathcal M_0=\begin{bmatrix}M_1& 0& -c_\beta s_W m_Z&s_\beta s_W m_Z\\
0& M_2& c_\beta c_W m_Z&-s_\beta c_W m_Z\\
-c_\beta s_W m_Z& c_\beta c_W m_Z& 0&-\mu\\
s_\beta s_W m_Z& -s_\beta c_W m_Z& -\mu&0\\
\end{bmatrix},\  
\end{equation}
\be
 \mathcal M^\pm=\begin{bmatrix}0 & X^T\\X&0
\end{bmatrix},\quad
X=\begin{bmatrix}M_2& \sqrt{2}s_\beta m_W\\ \sqrt{2}c_\beta m_W&\mu\\
\end{bmatrix}
\ee
where again all parameters are evaluated at the scale $\mathcal Q_0$,  $s_W=\sin \theta_W$ and so on.

One of the most appealing features of the MSSM, in the presence of $R$-parity, a symmetry preventing proton decay, is its capability to provide a viable DM candidate as the lightest supersymmetric particle (LSP): in particular the lightest neutralino $\chi_1^0$. However, from the plethora of DM searches in direct detection experiments, large regions of the parameter space of neutralino DM have been excluded~\cite{Aprile:2018dbl}. One scenario still alive is a nearly pure Higgsino with a mass $\sim$ 1.1 TeV~\cite{Roszkowski:2017nbc,Kowalska:2018toh}. This happens whenever $|\mu|\ll M_1,M_2$ and $|\mu|\simeq 1.1$ TeV. In this case the coupling of $\chi_1^0$ with the proton comes at tree-level from the coupling with the Higgs $h$, $\bar\chi_1^0\chi_1^0 h$, induced by the mixing of Higgsinos with gauginos, leading to spin-independent cross-sections with heavy nuclei. The spin-independent cross-section with the proton is bounded by the XENON1T experiment~\cite{Aprile:2018dbl} which yields, for $m_{\chi_1^0}\simeq 1$ TeV, the 90 \% C.L. bound $\sigma^{\rm SI}_p\lesssim 9\times 10^{-10}$ pb, which we will hereafter consider as a conservative limit.
\begin{figure}[htb]
\centering
\includegraphics[width=7.cm]{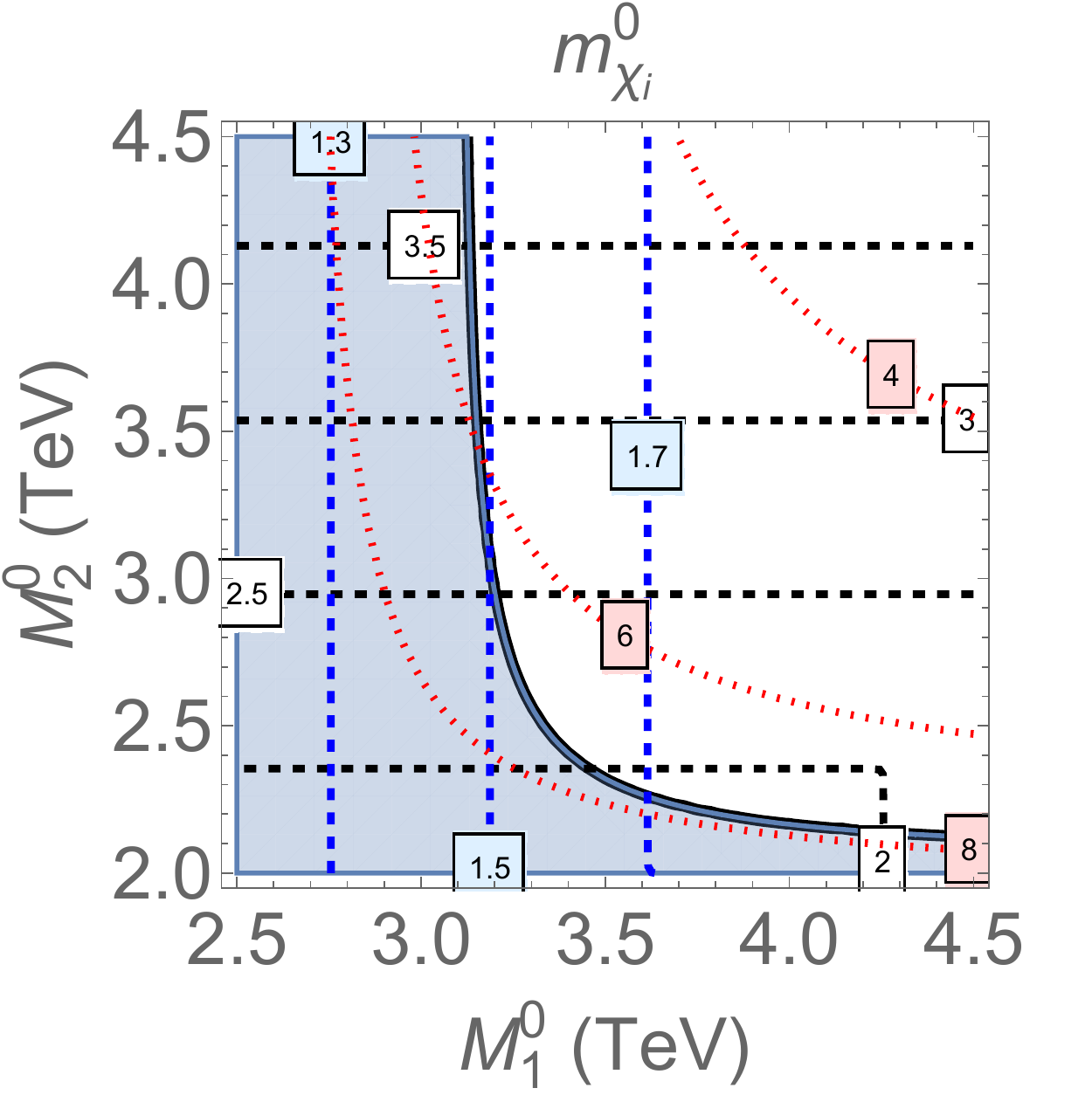} \hspace{0.5cm}
\includegraphics[width=7.7cm]{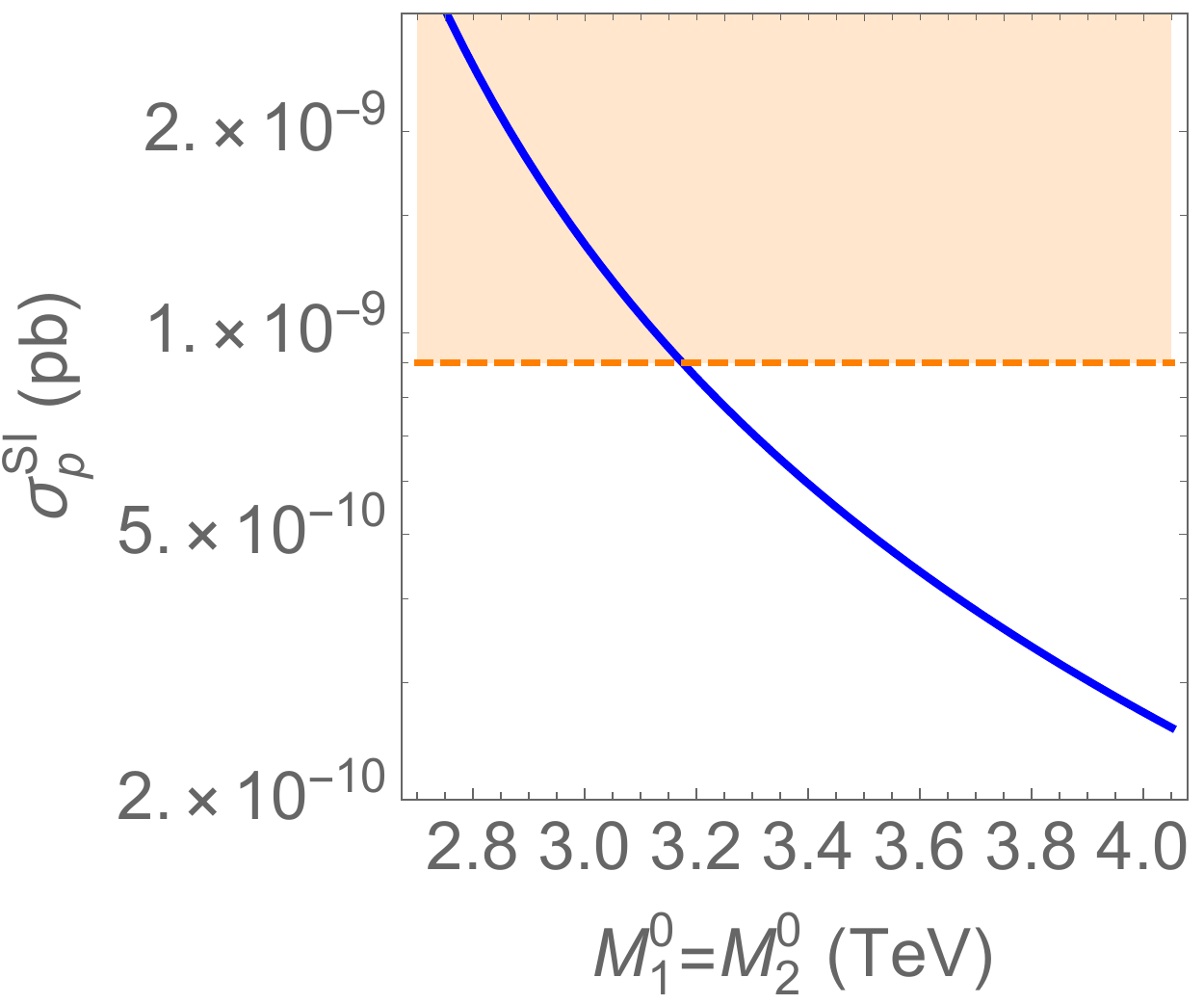} 
\caption{\it Left panel: The shadowed region is forbidden by the spin-independent cross-section for $\bar\chi_1^0\chi_1^0$-proton in the plane $(M_1^0,M_2^0)$. Vertical (horizontal) dashed lines are mass values of $\chi^0_3$ ($\chi^0_4$)  in TeV. Dotted lines are $m_{\chi_2^0}-m_{\chi_1^0}$ in GeV. Right panel: Plot of $\sigma_{p}^{\rm SI}$ (in pb) as a function of $M_1^0=M_2^0$ in TeV.}
\label{fig:SI}
\end{figure} 

In Fig.~\ref{fig:SI} (left panel) we show the (thick solid) contour line of  $\sigma^{SI}_p=9\times 10^{-10}$ pb in the plane of Majorana masses $(M_1^0,M_2^0)$. Thus the shadowed region is forbidden by the XENON1T experiment~\cite{Aprile:2018dbl}. The mass spectrum is then: i) The LSP $\chi_1^0$ with a mass $\sim 1.1$ TeV in all the shown region, while the next to lightest neutralino $\chi_2^0$, has a mass $m_{\chi_2^0}$  larger than $m_{\chi_1^0}$ by a few GeV, as we can see from the dotted contour lines. ii) The heavy states $\chi_3^0$ ($\chi_4^0$), with mass labels in TeV, are the vertical (horizontal) dashed lines. Therefore we can see that $\chi_1^0$ and $\chi_2^0$ are almost degenerate in mass around 1.1 TeV, while $m_{\chi_3}$ depends mainly on $M^0_1$, and $m_{\chi_4^0}$ depends mainly on $M^0_2$. In the allowed region we infer that $M^0_1\gtrsim 3$ TeV which corresponds to $m_{\chi_3^0}\gtrsim 1.5$ TeV, while if we want to stick to the lowest possible values of $m_{\chi_3^0}$ we need to consider the region where $M^0_2\gtrsim M_1^0$, although the larger $M_2^0$ the larger $m_{\chi_4^0}$. We conclude from this analysis that, in order to obtain the lightest possible neutralino spectrum, we should consider that $M_1^0\sim M_2^0$. As $M_a$ ($a=1,2,3$) evolves with the RGE as the couplings $\alpha_a$, it is a sensible condition to consider the unification condition $M_1^0=M_2^0$, as we will do hereafter.
\begin{figure}[htb!]
\centering
\includegraphics[width=7cm]{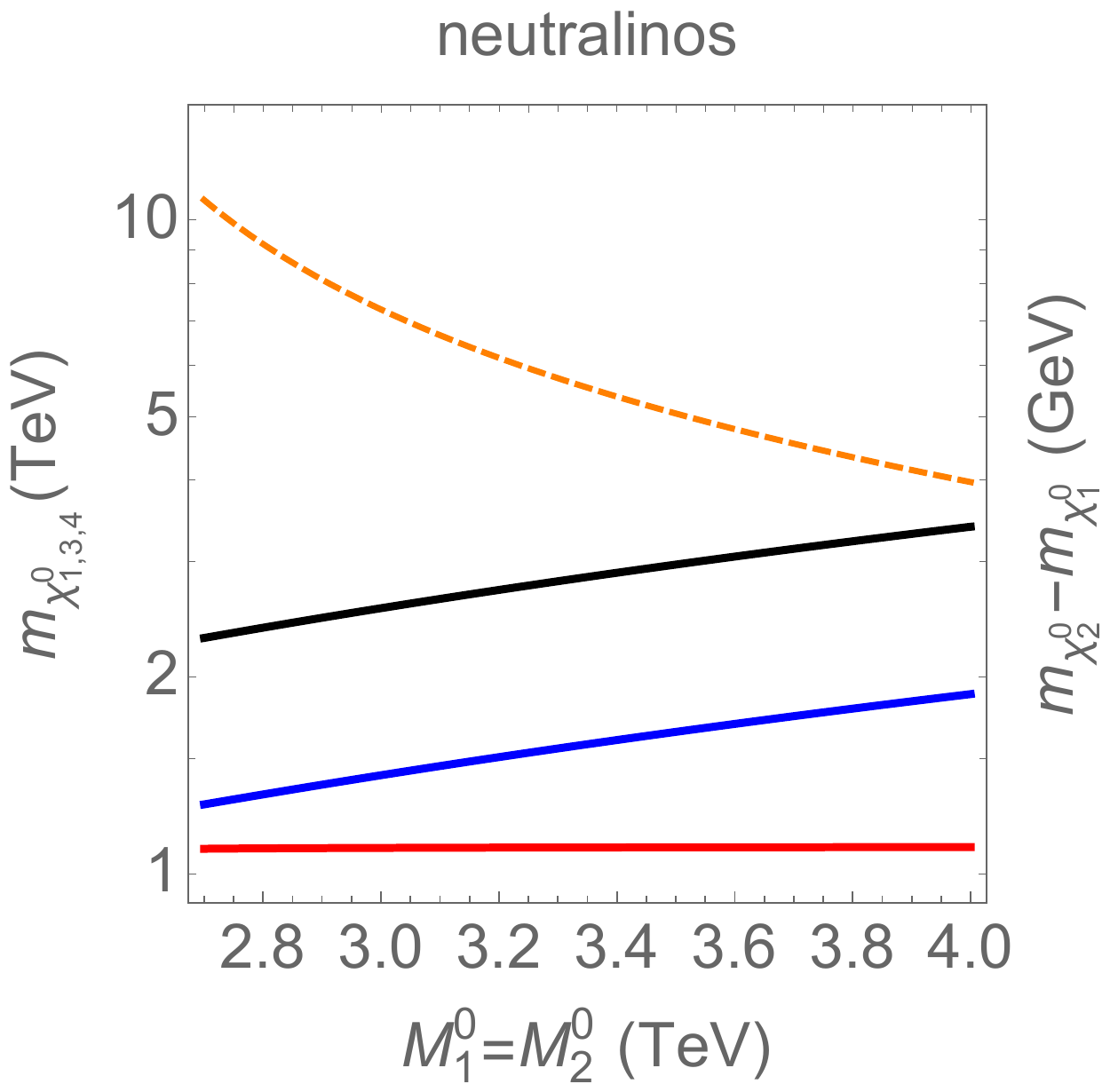} \hspace{0.5cm}
\includegraphics[width=7.cm]{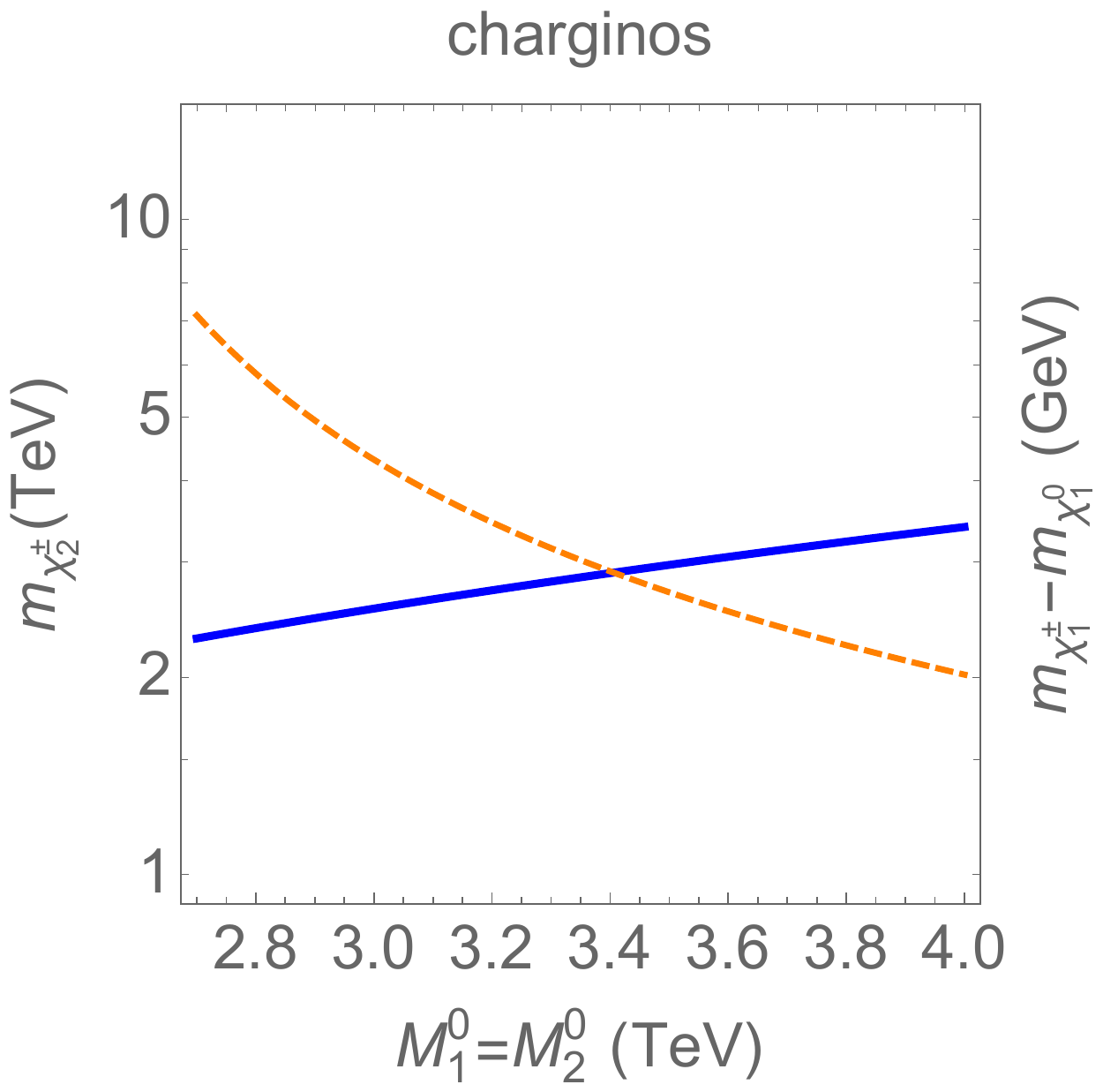} 
\caption{\it Mass spectra (in TeV) for neutralinos (left panel) and charginos (right panel) in solid lines for $M_1^0=M_2^0$ and $\mu=1.1$ TeV. The mass differences $m_{\chi_2^0}-m_{\chi_1^0}$ (left panel) and $m_{\chi_1^\pm}-m_{\chi_1^0}$ (right panel) are in GeV units.}
\label{fig:masses-inos}
\end{figure} 

In the right panel of Fig.~\ref{fig:SI} we consider the case where $M_1^0=M_2^0$ and plot the spin independent cross-section $\sigma_p^{\rm SI}$ as a function of $M_1^0=M_2^0$. We also plot the exclusion region from the XENON1T experimental results which translates into the lower bound $M_1^0=M_2^0\gtrsim 3.2$ TeV at 90\% C.L. The mass spectra for neutralinos (charginos) are plotted in the left (right) panel of Fig.~\ref{fig:masses-inos} in solid lines, while the mass differences $m_{\chi_2^0}-m_{\chi_1^0}$ (left panel) and $m_{\chi_1^\pm}-m_{\chi_1^0}$ (right panel) in GeV are in dashed lines. Using the direct detection cross section bound from the right panel of Fig.~\ref{fig:SI} we obtain the following restrictions on the neutralino and chargino running masses:
\begin{align}
&m_{\chi_1^0}\simeq1.1\ \textrm{TeV},\quad m_{\chi_3^0}\gtrsim 1.5\ \textrm{TeV},\quad  m_{\chi_4^0}\gtrsim 2.7\ \textrm{TeV},\quad  m_{\chi_2^0}- m_{\chi_1^0}\lesssim 6.3\ \textrm{GeV} \nonumber\\
& m_{\chi_2^\pm}\gtrsim 2.7\ \textrm{TeV},\quad   m_{\chi_1^\pm}- m_{\chi_1^0}\lesssim 3.5\ \textrm{GeV}
\label{eq:bounds}
\end{align}

In the following, in order to minimize the neutralino and chargino mass spectra, we will consider the benchmark case defined by
\be
M_1^0=M_2^0\simeq 3|\mu|\simeq 3.3\ \textrm{TeV}
\label{eq:m12}
\ee
which generates the spectrum given by the lower bounds in Eq.~(\ref{eq:bounds}). Larger values of $M_1^0=M_2^0$ could be equally well considered but they would lead to heavier spectra.

\section{Scenarios of Supersymmetry Breaking}
\label{sec:susybreaking}

At the low scale $\mathcal Q_0$, the soft-breaking masses $m_Q$ and $m_U$, and the mixing term $A_t$, can be obtained by integrating the RGE, and have been fitted in Ref.~\cite{Delgado:2014kqa}. After imposing the EoM, Eqs.~(\ref{eq:EoM}), one gets
\begin{align}
m_Q^2&=(m_Q^0)^2+\sum_a d_a f_a\cdot (M_a^0)^2+\frac{1}{3}F,\quad (d_1,d_2,d_3)=\left(-\frac{1}{15},1,\frac{8}{3}\right)
\label{eq:mQ}
\\
m_U^2&=(m_U^0)^2+\sum_a c_a f_a \cdot(M_a^0)^2+\frac{2}{3}F,\quad (c_1,c_2,c_3)=\left(\frac{1}{3},-1,\frac{8}{3}\right)
\label{eq:mU}
\end{align}
where the functions $F$  and $f_a$ are defined by
\begin{align}
F&=-(m_{H_U}^0)^2-\left(\mu^2+\frac{1}{2}m_Z^2\right)\frac{t_\beta^2-1}{t_\beta^2}+\frac{m_{H_D}^2}{t_\beta^2}
\label{eq:f}\\
f_a&=\frac{1}{b_a}\frac{\alpha_a^2(M)-\alpha_a^2(\mathcal Q_0)}{\alpha_a^2(M)},\quad (b_1,b_2,b_3)=\left( \frac{33}{5},1,-3 \right)
\label{eq:fa}
\end{align}
where $m_{H_D}$ is given in Eq.~(\ref{eq:mHD}).
In the same way the mixing parameter at the low scale can be written as
\be
A_t=\sum_a \gamma_a M_a^0+\gamma_A A_t^0
\label{eq:At}
\ee
where the coefficients $\gamma_a$ and $\gamma_A$ are determined numerically, and fitted in Ref.~\cite{Delgado:2014kqa}.

In this paper we want to find the region of parameters in Eq.~(\ref{eq:BC}) consistent with the condition of a nearly pure Higgsino, with a mass at $\sim$1.1 TeV, being the LSP and a good DM candidate, and satisfying the EoM of Eq.~(\ref{eq:EoM}), and all present experimental constraints. We will concentrate in two general models: i) The case of Universal Higgs Masses (UHM), a very popular model inspired by supergravity/superstring constructions, also dubbed as constrained MSSM (CMSSM) and, ii) A model where we make a separation of soft breaking masses in the Higgs and sfermion sectors, which is motivated by string constructions if both sectors are differently located in the higher dimensional compact space, which is dubbed as Non-Universal Higgs Masses (NUHM)~\cite{AbdusSalam:2011fc}. In both cases we will separately consider the case where all Majorana gaugino mass are unified at the high scale, and the case of non-universal gaugino masses~\cite{Kaminska:2013mya}, where we will concentrate in the phenomenologically interesting case where only the gluino mass does not unify with the electroweakino masses at the high scale $M$, and such that the gluino is on the verge of experimental detection at the LHC. 

\subsection{Universal Higgs Masses}

In this section we will assume the case of UHM so that the boundary conditions for the scalar sector are
\be
m_Q^0=m_U^0=m_D^0=m_ L^0=m_E^0=m_{H_U}^0=m_{H_D}^0\equiv m_0,
\ee
where only the third generation squark masses are relevant for our study. 

\subsubsection{Universal Gaugino Masses}
As for the Majorana gaugino masses we will first assume the case of universal gaugino masses, i.e.~Majorana masses which unify at the high (unification) scale $M$, as
\be
M_1^0=M_2^0=M_3^0\equiv m_{1/2}
\label{eq:UGM}
\ee
in which case we have as free parameters in our model $(m_0,m_{1/2},A_t^0,t_\beta)$, once we have fixed $\mu\simeq 1.1$ TeV. Moreover, as we want to minimize as much as possible the mass of charginos and neutralinos, we will adopt the value 
\be
m_{1/2}=3\mu
\ee
such that the mass spectrum of neutralinos and charginos is essentially given by the lower bounds in Eq.~(\ref{eq:bounds}), while the gluino running mass at the low scale $\mathcal Q_0$ is $M_{\tilde g}\simeq 6.7$ TeV, out of reach of LHC experimental searches.
\begin{figure}[htb!]
\centering
\includegraphics[width=7cm]{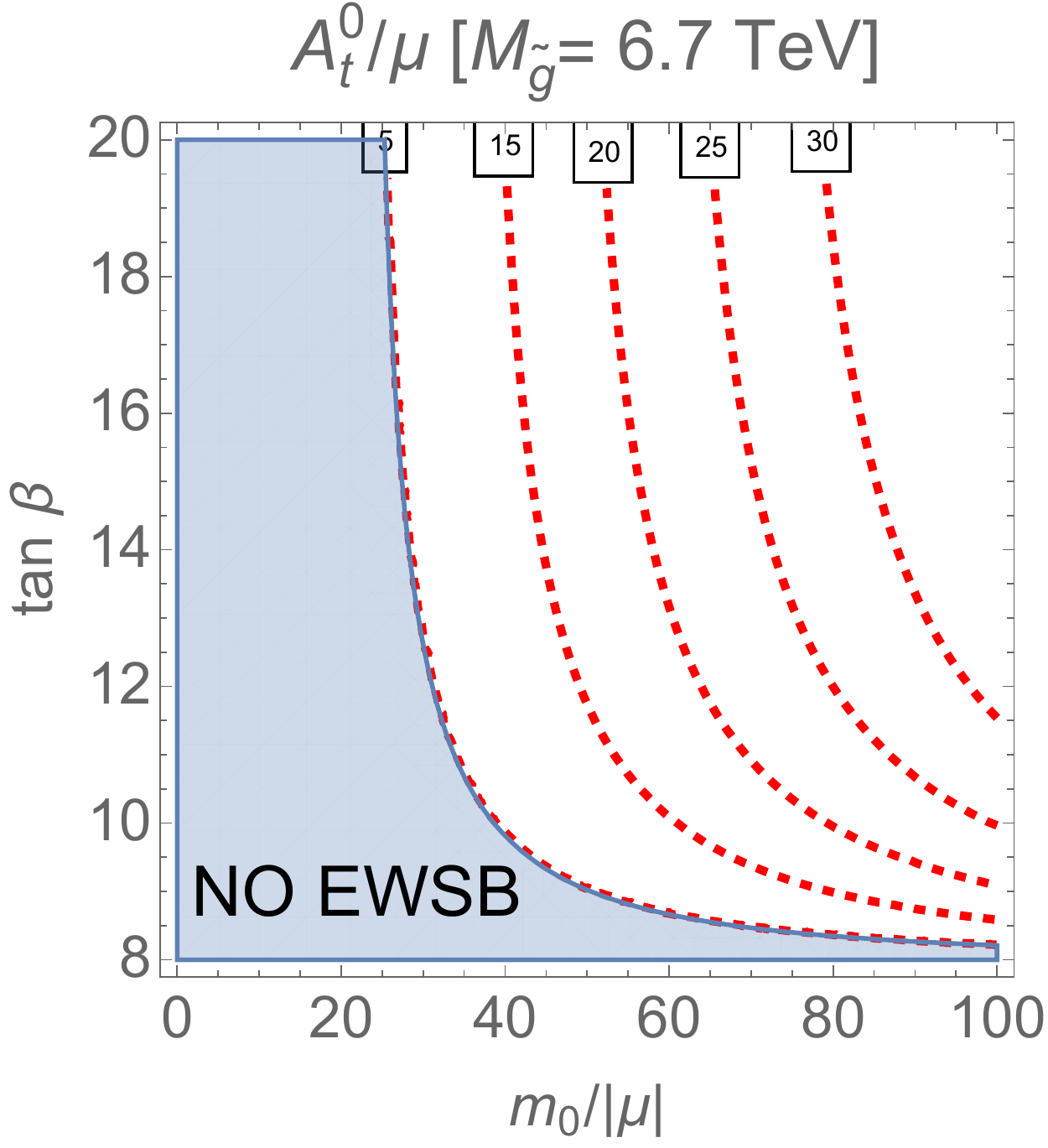} \hspace{0.5cm}
\includegraphics[width=7.cm]{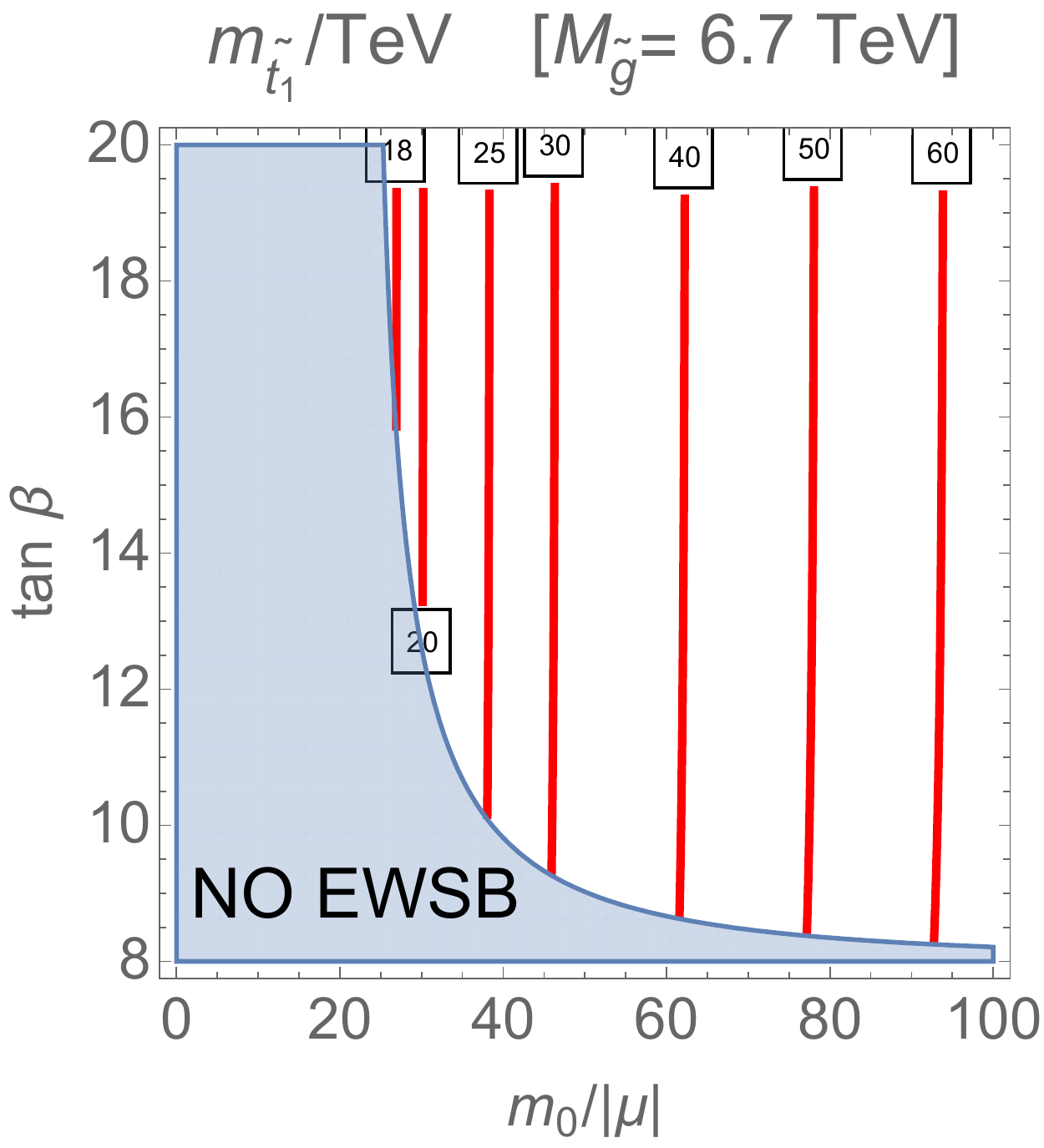} 
\caption{\it Left panel: Contour (dashed) lines of $A_t^0/\mu$ in the plane $(m_0/|\mu|,t_\beta)$ satisfying the EoM of Eq.~(\ref{eq:EoM}) for universal gaugino masses. The shadowed region is forbidden by the EWSB condition. Right panel: The same for contour (solid) lines of the lightest stop running mass $m_{\tilde t_1}$ in TeV units.}
\label{fig:CMSSM}
\end{figure} 
The remaining three parameters $(m_0,A_t^0,t_\beta)$ have to satisfy the EoM, Eq.~(\ref{eq:EoM}). The result is shown in the left panel of Fig.~\ref{fig:CMSSM} where we exhibit contour lines of $A_t^ 0/\mu$ (red dashed lines) in the plane $(m_0/|\mu|,t_\beta)$. The shadowed region corresponds to values of the parameters where the EoM is not satisfied for real values of the parameters, and thus there is no EWSB. 

As we can see, the EWSB condition translates into an absolute lower bound on the parameter $t_\beta$ as,
$
t_\beta\gtrsim 8
$. Moreover, using the expressions for $m_Q$, $m_U$ and $A_t$ in Eqs.~(\ref{eq:mQ}), (\ref{eq:mU}) and (\ref{eq:At}), respectively, one can easily compute the (running) mass spectrum for stops $(\tilde t_1, \tilde t_2)$, where we are using the convention that $\tilde t_1$ is the lightest stop. Contour lines of $m_{\tilde t_1}$ are provided (red solid lines) in the right panel of Fig.~\ref{fig:CMSSM}, from where we see that the solution with light stops is prevented by the EoM. In fact we see that the EWSB condition implies the lower bound $m_{\tilde t_1}\gtrsim 18$ TeV. As a consequence this scenario predicts superheavy masses in the sfermion and gluino sectors, completely out of reach of future searches at the LHC. Furthermore,
as the values of the mixing parameters are tiny compared mainly with the values of stop masses,  we can conclude that the mixing parameter at the low scale $\mathcal Q_0$ is negligible compared with the values of the relevant supersymmetric masses~\footnote{In particular we get that in the considered range $|A_t-\mu/t_\beta|<0.1\sqrt{m_{\tilde t_1}m_{\tilde t_2}}$.}. Due to the heavy spectrum this scenario might be in tension with the correct value of the Higgs mass (see Ref.~\cite{Draper:2013oza} for more details).

\subsubsection{Non-Universal Gaugino Masses}
As we have seen that, for universal gaugino masses, the gluino is heavy and out of reach of LHC experimental searches, we will now explore a scenario where the gluino is on the verge of experimental detection, which can be done if the gluino Majorana mass is different at the high scale $M$ from the bino and wino Majorana masses. In particular we will assume Eq.~(\ref{eq:m12}) for $M_{1,2}^0$, and a value for $M_3^0$ corresponding to a running gluino mass, at the low scale, $\mathcal Q_0$ of 2.2 TeV, i.e.
\be
M_1^0=M_2^0=3\mu,\quad M_{\tilde g}=2.2 \textrm{ TeV} ,
\label{eq:NUGM}
\ee
in which case the free parameters of the model are still $(m_0,A_t^0,t_\beta)$ as in the previous case of universal gaugino masses, but of course with different realization of the EWSB conditions in the gaugino sector.
\begin{figure}[htb!]
\centering
\includegraphics[width=7cm]{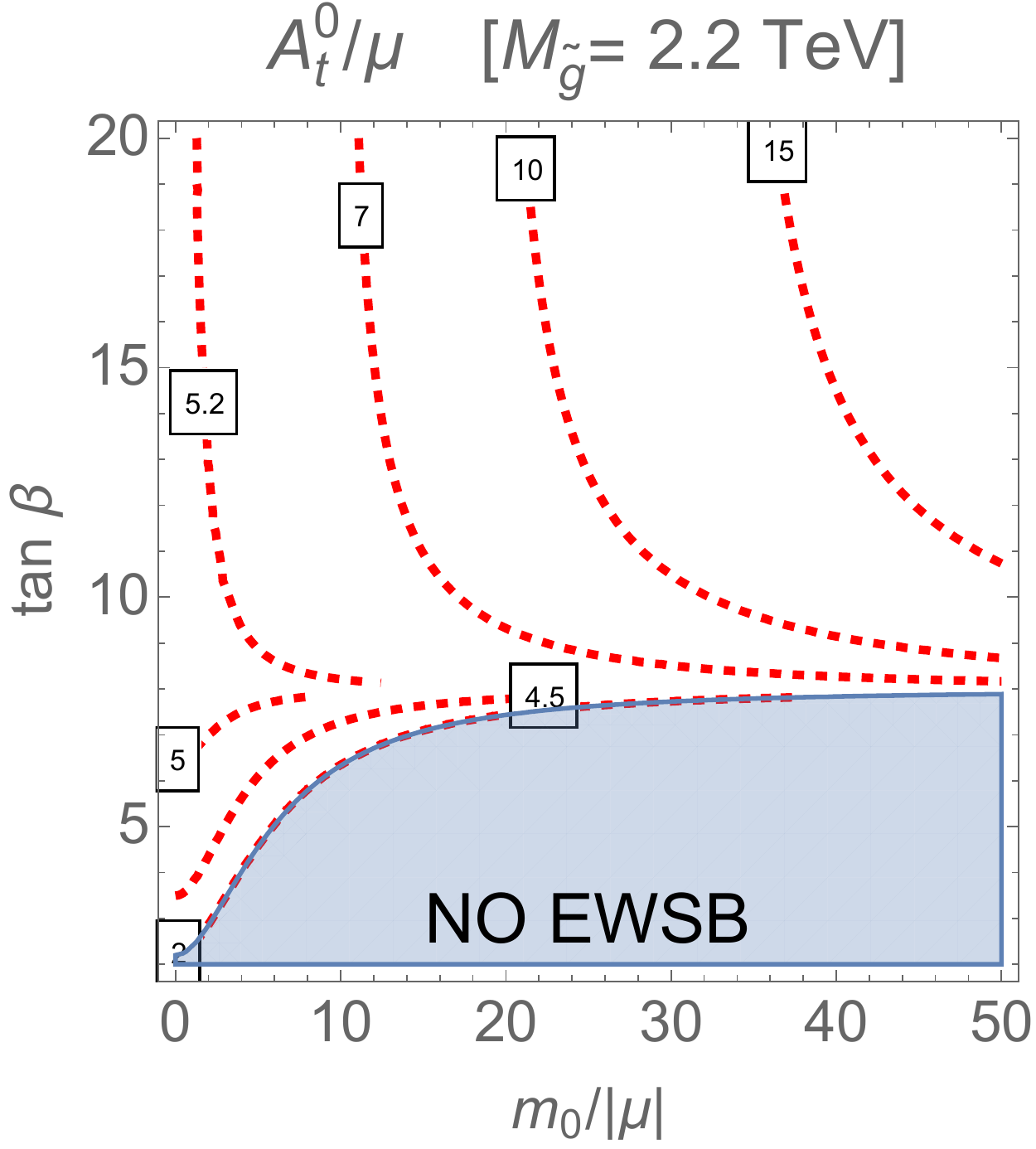} \hspace{0.5cm}
\includegraphics[width=7.cm]{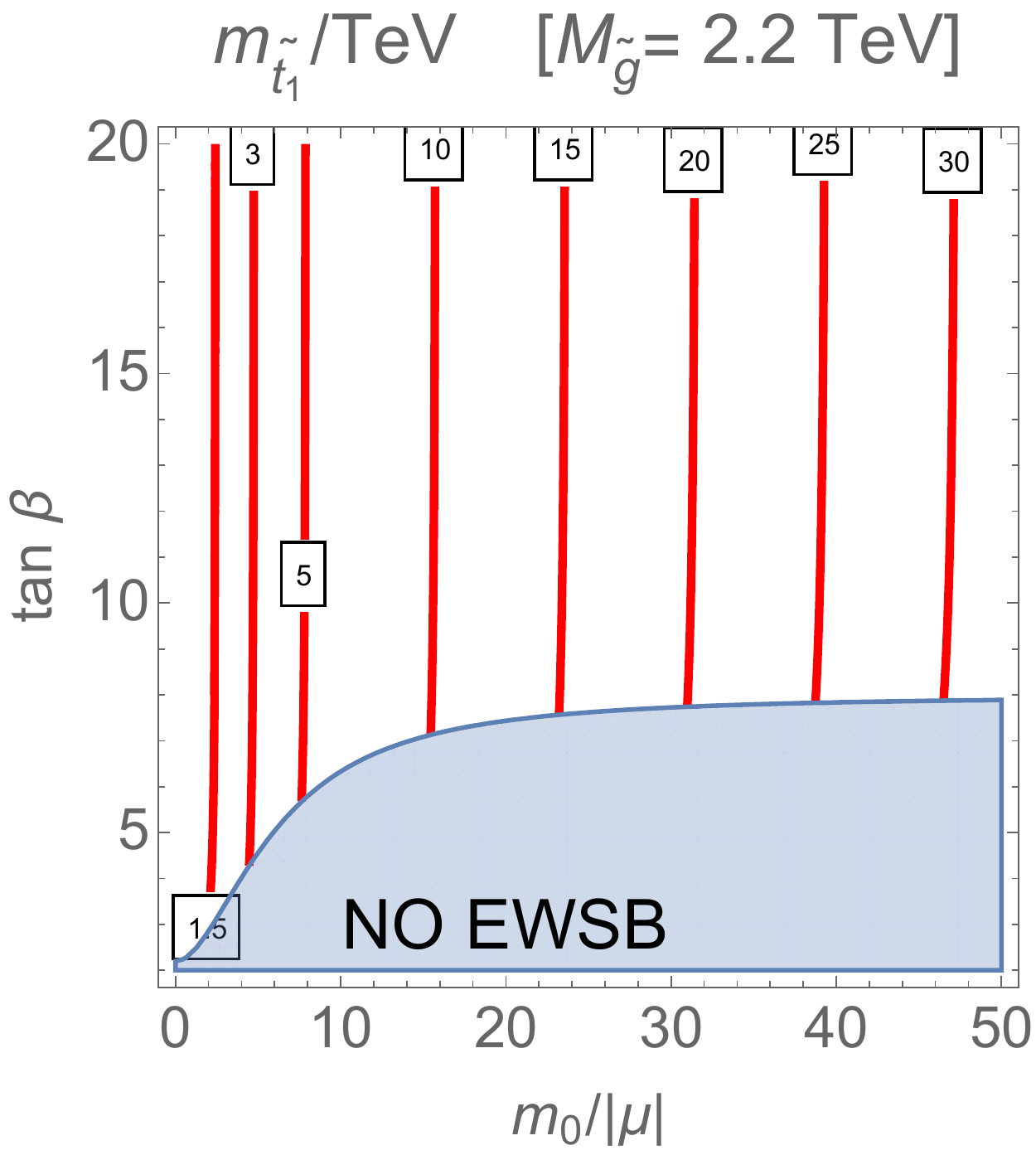} 
\caption{\it Left panel: Contour (dashed) lines of $A_t^0/\mu$ in the plane $(m_0/|\mu|,t_\beta)$ satisfying the EoM of Eq.~(\ref{eq:EoM}) for non-universal gaugino masses such that $M_{\tilde g}=2.2$ TeV. The shadowed region is forbidden by the EWSB condition. Right panel: The same for contour (solid) lines of the lightest stop running mass $m_{\tilde t_1}$ in TeV units.}
\label{fig:CMSSM-NUGM}
\end{figure} 
%

After imposing the EoM (\ref{eq:EoM}) the values of the parameters are provided in the left panel of Fig.~\ref{fig:CMSSM-NUGM}, which shows (dashed) contour lines of $A_t^0/\mu$ in the plane $(m_0/|\mu|,t_\beta)$. We see that for large values of $m_0$ there is still the lower bound $t_\beta\gtrsim 8$, while there is no bound on $t_\beta$ for small values of $m_0$. Similarly (solid) contour lines of $m_{\tilde t_1}$ in TeV units are shown in the right panel of Fig.~\ref{fig:CMSSM-NUGM}. Moreover we see, from the right panel of Fig.~\ref{fig:CMSSM-NUGM}, that there is no lower bound on the value of $m_{\tilde t_1}$, as there is no experimental bound on $m_{\tilde t_1}$ for the value of the LSP mass, $m_{\tilde\chi_1^0}\simeq 1.1$ TeV~\cite{Aad:2020sgw}~\footnote{The only constraint in the present scenario would be of course imposing the lightest neutralino to be the LSP, i.e.~$m_{\tilde t_1}>m_{\tilde\chi_1^0}$.}. However we have found that in all the region of the parameter space the mixing in the stop sector is small, $|A_t-\mu/t_\beta|\lesssim 0.2 \sqrt{m_{\tilde t_1}m_{\tilde t_2}}$ so that, in the light stop region (small values of $m_0$), the scenario would fail to describe the correct value of the Higgs mass. The successful region to accommodate the Higgs mass would require to go to large values of $t_\beta$ and large values of the lightest stop masses (say $15\lesssim m_{0}/|\mu|\lesssim 25$) in which case all sfermions would be out of reach of the experimental LHC searches, and only the gluino could be discovered in the near future.  A benchmark case is provided in Tab.~\ref{tab:MSSM}, where we have chosen $t_\beta=10$ and $m_0=12\mu$, with heavy spectrum of supersymmetric scalars and heavy Higgs sector, and where, of course, only the gluino could be detectable at the LHC. The last column contains the prediction for the light Higgs mass $m_h$, for which we have used FeynHiggs from Refs.~\cite{Heinemeyer:1998yj,Heinemeyer:1998np,Degrassi:2002fi,Frank:2006yh,Hahn:2013ria,Bahl:2016brp,Bahl:2017aev,Bahl:2018qog}, with an estimated theoretical error $\Delta m_h\lesssim 1$ GeV~\cite{Bahl:2019hmm}.
\begin{table}
\vspace{0.5cm}
\centering
\begin{tabular}{||c||c|c|c|c|c|c|c|c|c|c||}
\hline\hline
Field & $\tilde t_1$ & $\tilde t_2$ & $\tilde b_L,\ \tilde Q_L^i$ &$\tilde u_R^i$ & $\tilde d_R^a$& $\tilde \ell^a_L$ & $\tilde e_R^a$& $H^{0,\pm},A$ &$\tilde g$ &$h$
\\ \hline
Mass (TeV)& 7.7 & 11.1 & 13.5 & 13.4  &13.3 & 13.4& 13.3& 13.4& 2.2 & 124\\
\hline\hline
\end{tabular}
\caption{\it Benchmark supersymmetric spectrum for values of the parameters: $t_\beta=10$, and $m_0\simeq 12\mu$. All masses are in TeV units, except the SM Higgs ($h$) mass which is in GeV. Generation indices run as: $a=1,2,3$, $i=1,2$. $SU(2)_L$ doublets are indicated by $\tilde Q_L$ and $\tilde\ell_L$, for squarks and sleptons, respectively.}
\label{tab:MSSM}
\end{table}

\subsection{Non-Universal Higgs Masses}

In the previous section we have seen that in the case of UHM the only way of solving the EoM and describing the correct Higgs mass is  with a very heavy squark mass spectrum, out of reach of LHC searches. In this section we will show that one way of avoiding this feature is imposing different unified masses at the high unification scale $M$ for squarks and sleptons, and the Higgs sector. In particular we will introduce the boundary conditions at the scale $M$ given by
\be
m_Q^0=m_U^0=m_D^0=m_ L^0=m_E^0\equiv m_0,\quad m_{H_U}^0=m_{H_D}^0\equiv m_H
\ee
 \subsubsection{Universal Gaugino Masses}
We will first consider the case of universal gaugino masses at the scale $M$ given by the boundary conditions in Eq.~(\ref{eq:UGM}). In this case the free parameters are $(m_0,m_H,A_t^0,t_\beta)$. Motivated by the purpose of describing the Higgs mass with a light spectrum, as much as possible, we will consider the case of large $t_\beta$, and fix $t_\beta=10$, which is large enough to contribute efficiently to the tree-level calculation of the Higgs mass and small enough to consistently allow the neglect of the bottom Yukawa coupling, as we are doing in our analytical computation. %
\begin{figure}[h!]
\centering
\includegraphics[width=7cm]{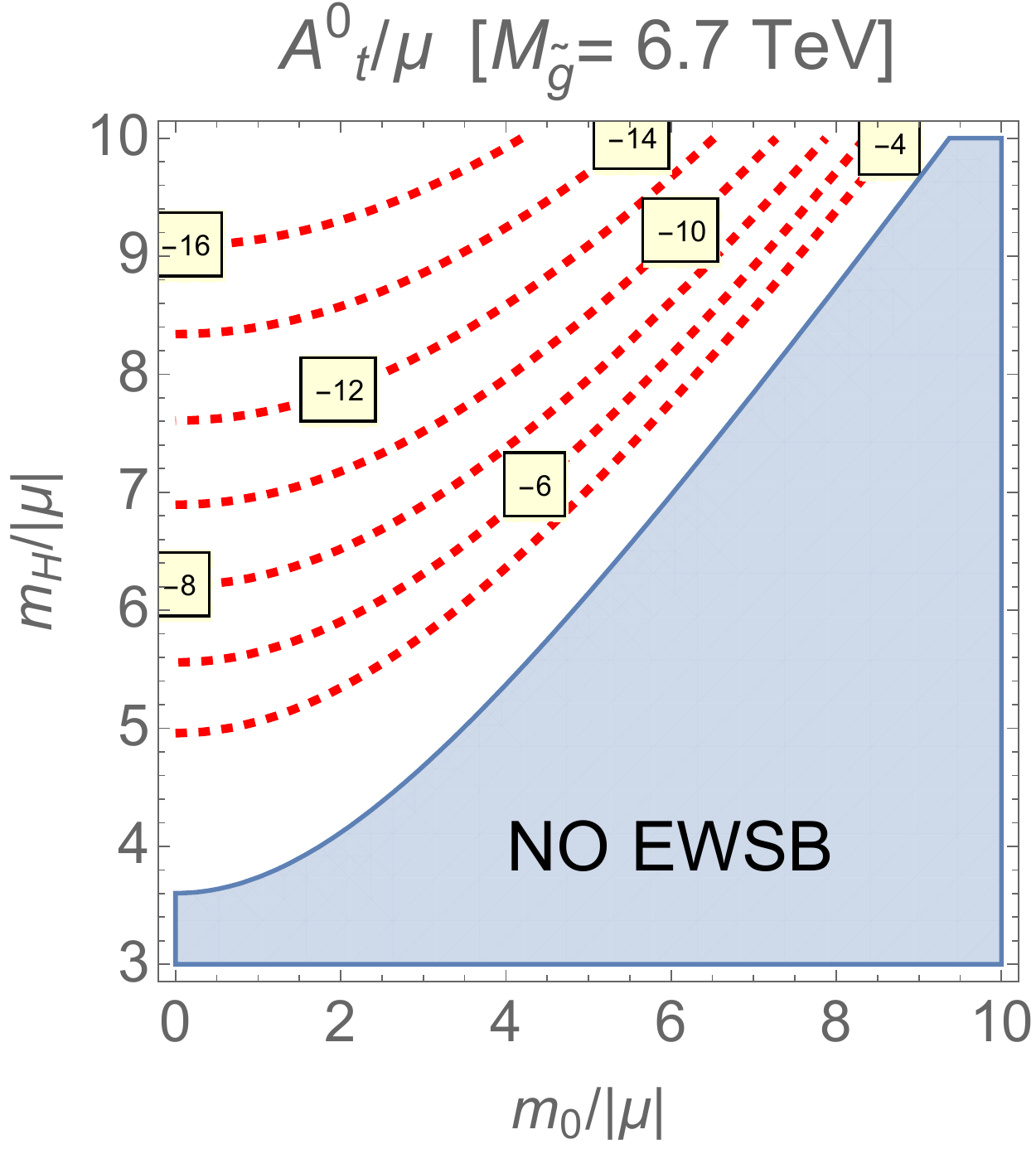} \hspace{0.5cm}
\includegraphics[width=7.cm]{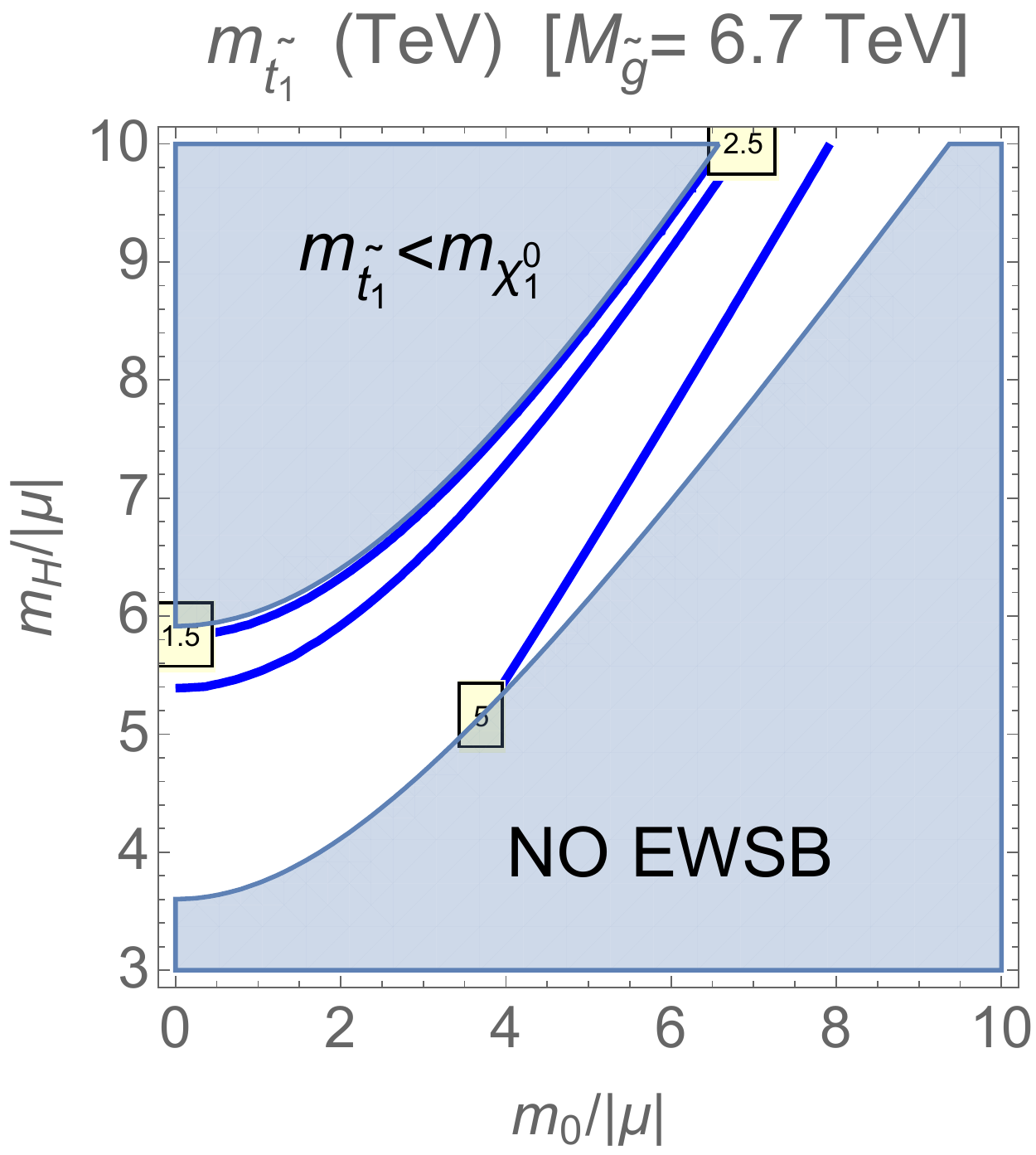} 
\caption{\it  Left panel: Contour (dashed) lines of $A_t^0/\mu$ in the plane $(m_0/|\mu|,m_H/|\mu|)$ for $t_\beta=10$ satisfying the EoM of Eq.~(\ref{eq:EoM}) for universal gaugino masses. The shadowed region is forbidden by the EWSB condition. Right panel: The same for contour (solid) lines of the lightest stop running mass $m_{\tilde t_1}$ in TeV units.}
\label{fig:HMSSM}
\end{figure} 
The resulting three parameters $(m_0,m_H,A_t^0)$ can be confronted with the EoM, Eq.~(\ref{eq:EoM}). 

In the left panel of Fig.~\ref{fig:HMSSM} we show (dashed) contour lines of $A_t^0/\mu$ in the plane $(m_0/|\mu|,m_H/|\mu|)$, while the right panel  shows contour lines of $m_{\tilde t_1}$ in TeV units. In both panels the lower shadowed region is the forbidden region where there is no EWSB. In the right panel the upper shadowed region is the forbidden region where $m_{\tilde t_1}<m_{\tilde\chi_1^0}$ and the lightest stop would be the LSP. In the latter case we would need a lighter DM candidate, a condition that we are not exploring in the present paper. 
As we can see in the right panel of Fig.~\ref{fig:HMSSM}, for small values of $m_0$ and values of $m_H$ near the upper shadowed region, the lightest stop could be at reach of the future LHC running, and there are also relatively light states for the rest of squarks and sleptons. Moreover, in this region the stop mixing parameter can be maximal, i.e.~$|A_t-\mu/t_\beta|\simeq \sqrt{6} \sqrt{m_{\tilde t_1}m_{\tilde t_2}}$ and the Higgs mass can be easily accommodated by stops in the TeV region. 

A benchmark model with maximal mixing is presented in Tab.~\ref{tab:HMSSM}, where we give the tree level masses, in TeV, for the different scalars, and we have skipped the tiny splitting generated by the EWSB contribution. We can see that the lightest scalar is the third-generation right-handed slepton with a mass $\sim 1.2$ TeV. The last column's prediction is obtained from FeynHiggs in Refs.~\cite{Heinemeyer:1998yj,Heinemeyer:1998np,Degrassi:2002fi,Frank:2006yh,Hahn:2013ria,Bahl:2016brp,Bahl:2017aev,Bahl:2018qog}, with an estimated theoretical error $\Delta m_h\lesssim 2$~GeV~\cite{Bahl:2019hmm}.

\begin{table}[htb]
\vspace{0.5cm}
\centering
\begin{tabular}{||c||c|c|c|c|c|c|c|c|c|c||}
\hline\hline
Field & $\tilde t_1$ & $\tilde t_2$ & $\tilde b_L,\ \tilde Q_L^i$ &$\tilde u_R^i$ & $\tilde d_R^a$& $\tilde \ell^a_L$ & $\tilde e_R^a$& $H^{0,\pm},A$ &$\tilde g$& $h$
\\ \hline
Mass (TeV)& 2.0 & 4.8 & 6.2 & 5.9  &5.8 & 2.2& 1.2& 6.6& 6.7 & 125\\
\hline\hline
\end{tabular}
\caption{\it Benchmark supersymmetric spectrum for values of the parameters: $t_\beta=10$, $m_0\simeq 0$ and $m_H\simeq 6.3\, \mu$. All masses are in TeV units, except the SM Higgs mass ($h$) which is in GeV. Generation indices run as: $a=1,2,3$, $i=1,2$. $SU(2)_L$ doublets are indicated by $\tilde Q_L$ and $\tilde\ell_L$, for squarks and sleptons, respectively.}
\label{tab:HMSSM}
\end{table} 

\subsubsection{Non-Universal Gaugino Masses}
Here we will consider the case of non-universal gaugino masses at the scale $M$ corresponding to the boundary conditions in Eq.~(\ref{eq:NUGM}), still with free parameters given by $(m_0,m_H,A_t^0,t_\beta)$. 
\begin{figure}[htb!]
\centering
\includegraphics[width=7cm]{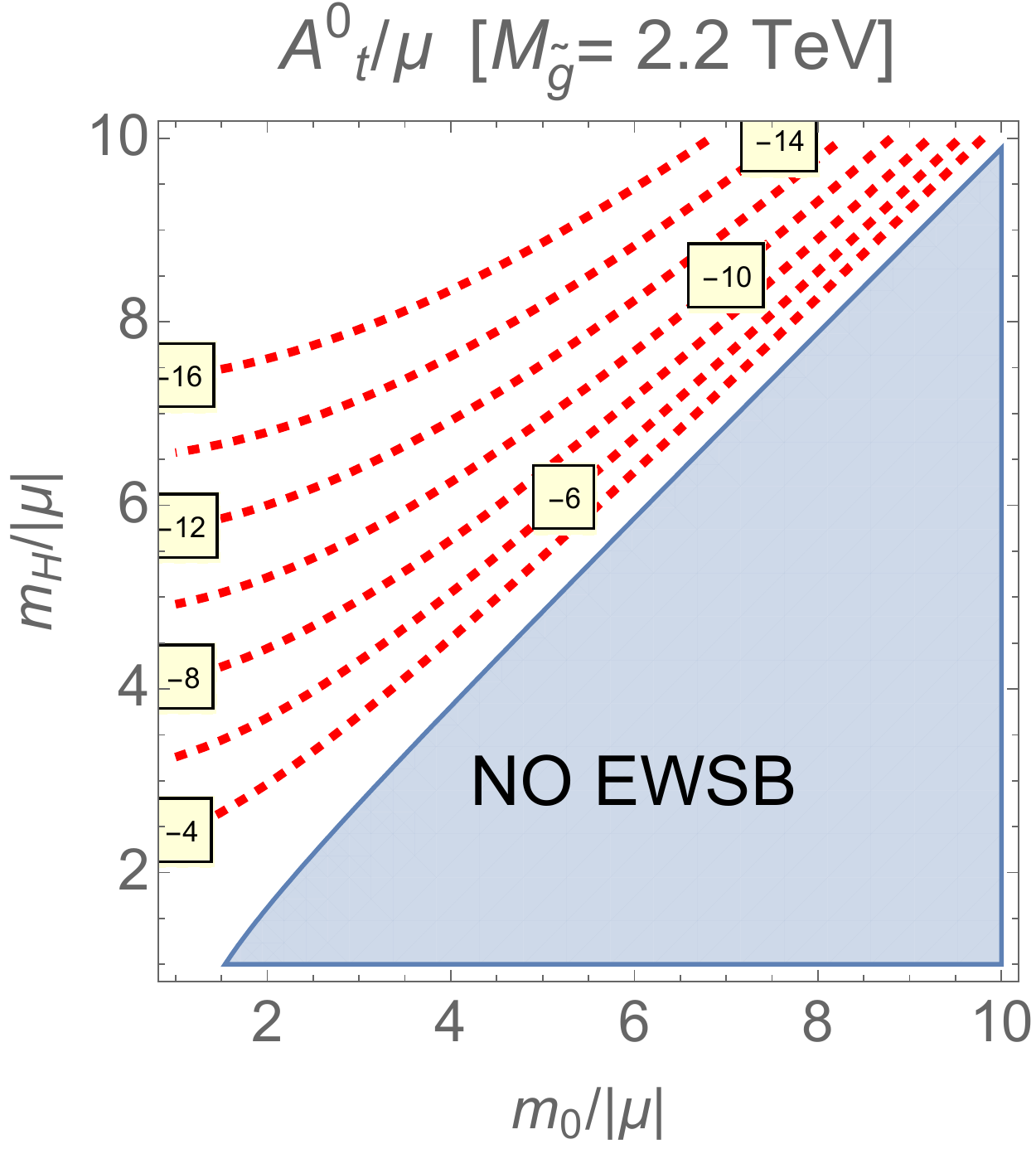} \hspace{0.5cm}
\includegraphics[width=7.cm]{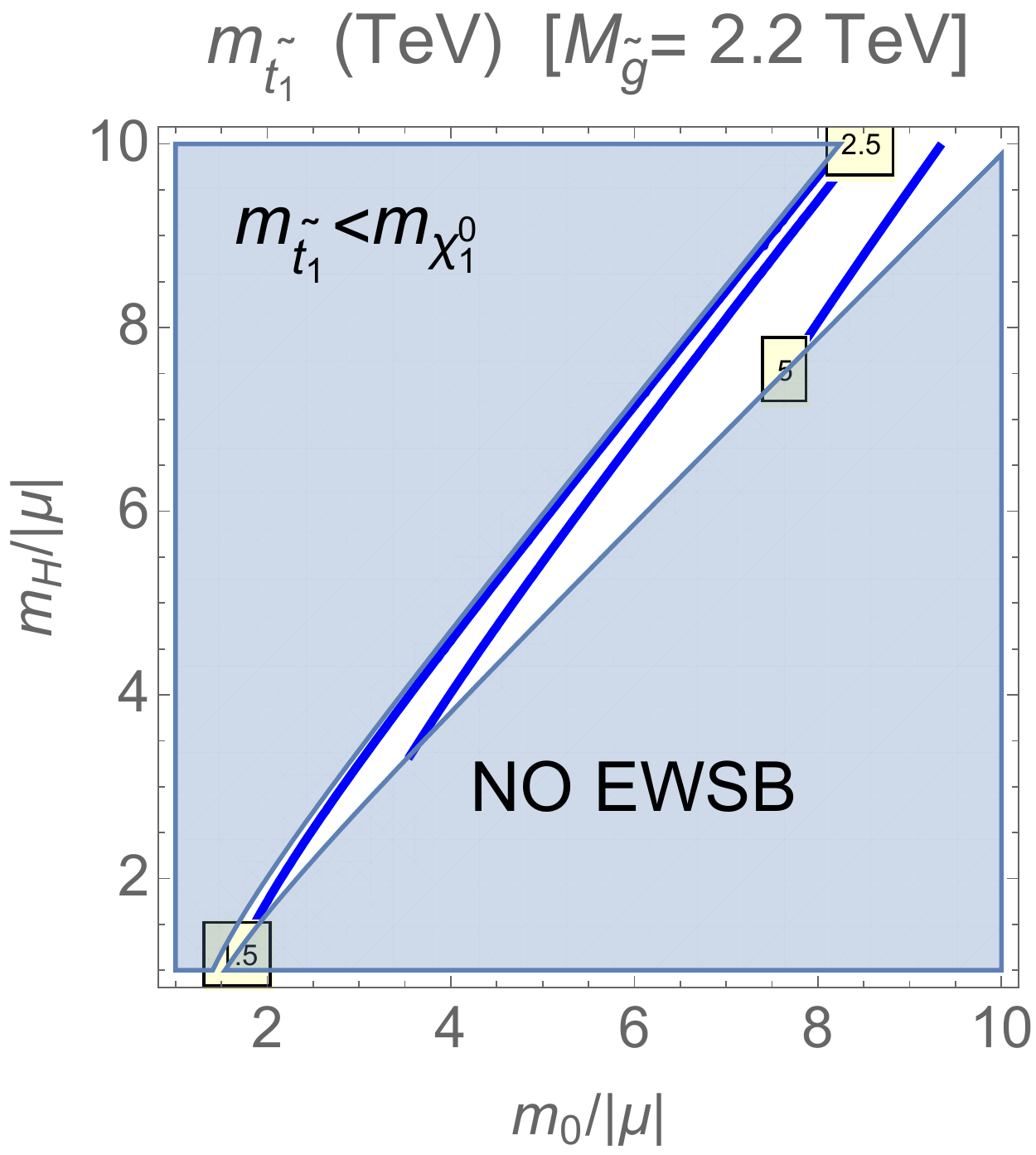} 
\caption{\it  Left panel: Contour (dashed) lines of $A_t^0/\mu$ in the plane $(m_0/|\mu|,m_H/|\mu|)$ for $t_\beta=10$ satisfying the EoM of Eq.~(\ref{eq:EoM}) for non-universal gaugino masses. The shadowed region is forbidden by the EWSB condition. Right panel: The same for contour (solid) lines of the lightest stop running mass $m_{\tilde t_1}$ in TeV units.}
\label{fig:HMSSM-NUGM}
\end{figure} 
As in the case of UHM, and mainly motivated by describing the correct value of the Higgs mass with a relatively light stop spectrum, we will fix $t_\beta=10$, so that the free parameters are $(m_0,m_H,A_t^0)$ that we will choose to satisfy  the EoM, Eq.~(\ref{eq:EoM}).

In the left panel of Fig.~\ref{fig:HMSSM-NUGM} we show (dashed) contour lines of $A_t^0/\mu$ in the plane $(m_0/|\mu|,m_H/|\mu|)$, and in the right panel we show contour lines of $m_{\tilde t_1}$ in TeV units. In both panels the lower shadowed region is the region where there is no EWSB. In the right panel the upper shadowed region is the region where $m_{\tilde t_1}<m_{\tilde\chi_1^0}$ and such that the lightest stop would be the LSP. 
\begin{table}
\vspace{0.5cm}
\centering
\begin{tabular}{||c||c|c|c|c|c|c|c|c|c|c||}
\hline\hline
Field & $\tilde t_1$ & $\tilde t_2$ & $\tilde b_L,\ \tilde Q_L^i$ &$\tilde u_R^i$ & $\tilde d_R^a$& $\tilde \ell^a_L$ & $\tilde e_R^a$& $H^{0,\pm},A$ &$\tilde g$ & $h$
\\ \hline
Mass (TeV)& 1.57 & 3 & 3.5 & 2.9  &2.8 & 3.1& 2.4& 2.8& 2.2 & 124\\
\hline\hline
\end{tabular}
\caption{\it Benchmark supersymmetric spectrum for the value of the gluino mass $M_{\tilde g}=2.2$ TeV, and values of the parameters: $t_\beta=10$, $m_0\simeq 1.9\,\mu$ and $m_H\simeq\, 1.5\mu$. All masses are in TeV units, except the SM Higgs mass ($h$) which is in GeV. Generation indices run as: $a=1,2,3$, $i=1,2$. $SU(2)_L$ doublets are indicated by $\tilde Q_L$ and $\tilde\ell_L$, for squarks and sleptons, respectively.}
\label{tab:HMSSM-NUGM}
\end{table} 
For the smallest possible values of $m_0$ and $m_H$ consistent with EWSB, $m_0\simeq 1.9| \mu|$ and $m_H\simeq 1.5 |\mu|$, the lightest stop is as light as possible within the present model (lighter than in the case of universal gaugino masses), and the rest of squarks and heavy Higgses are also lighter than in the case of universal gaugino masses because of a smaller renormalization from the gluino mass  and the smaller value of the common Higgs mass $m_H$, respectively, while sleptons are heavier, because of the larger value of the common masses $m_0$. In this region the stop mixing is near maximal, i.e.~$|A_t-\mu/t_\beta|\simeq \sqrt{6} \sqrt{m_{\tilde t_1}m_{\tilde t_2}}$, and the Higgs mass can be easily accommodated.

A benchmark model with maximal mixing is presented in Tab.~\ref{tab:HMSSM-NUGM}, where we give the tree level masses, in TeV, for the different scalars, and where, as done in Tab.~\ref{tab:HMSSM}, we have skipped the tiny splitting generated by the EWSB contribution. We can see that the lightest scalar is the lightest stop with a mass $\sim 1.6$ TeV. The last column's prediction is obtained from the code FeynHiggs, Refs.~\cite{Heinemeyer:1998yj,Heinemeyer:1998np,Degrassi:2002fi,Frank:2006yh,Hahn:2013ria,Bahl:2016brp,Bahl:2017aev,Bahl:2018qog}, with an estimated theoretical error $\Delta m_h\lesssim 2$ GeV~\cite{Bahl:2019hmm}.

\section{Experimental signatures}
\label{sec:phenomenology}

In this section we will comment on the different experimental signatures that the scenarios presented in this paper could have.
The common feature, and main motivation, for this analysis has been a Higgsino doublet with a mass of  1.1 TeV, and with splitting among the lightest components of several GeV, as can be seen in Eq.~(\ref{eq:bounds}). The production cross-section of such a Higgsino is too low for discovery at the LHC, but it can be produced at a 100 TeV collider, where the signal cross section is higher and it may be possible to create sufficient amounts of highly boosted charginos and neutralinos for discovery~\cite{Bramante:2014tba,Bramante:2014tba2, Low:2014cba}.

A better option for discovering the Higgsino LSP is via dark matter direct detection experiments~\cite{Bramante:2014tba,Bramante:2014tba2}. The detection prospects strongly depend on the bino/wino admixture in the LSP, as that admixture controls the strength of the LSP-LSP-Higgs vertex that drives spin-independent scattering rate off nuclei~\footnote{The LSP Higgsino can also be detected via its spin-dependent scattering off nuclei, although the prospects there are not as good~\cite{Kowalska:2018toh}.}. The LSP for our benchmark points is around $99\%$ pure Higgsino  --  a result of the large wino/bino mass --  so the spin-independent nuclear cross section for the benchmark range is of the order of few $10^{-10}$ pb~\cite{Kowalska:2018toh}; thus, the whole range escapes the current limit from XENON-1T~\cite{Aprile:2018dbl}. However, as shown in Ref.~\cite{Kowalska:2018toh}, an LSP Higgsino of this purity will be accessed in the next generation experiments, like XENON-nT or LZ~\cite{Mount:2017qzi}.  

Colored particles have higher cross sections and therefore can be easily produced at the LHC. In particular, the gluino is the one having the largest cross sections, being a QCD octet. In scenarios of universal gaugino masses, they sit at a mass around, or larger than, 6.7 TeV, which is completely out of reach for the LHC. On the other hand, when one deviates from universal boundary conditions, the gluino mass can be almost a free parameter, and we have decided to put it at 2.2 TeV, which is the current LHC bound~\cite{ATLAS:2018yhd}. One feature of all spectra presented here is that there exists a heavy neutralino, mostly bino, $\chi^0_3$, whose mass is in between the gluino and the Higgsino (LSP) masses; this fact will make the gluino decay, either directly to the LSP, or to  $\chi^0_3$ which will then, subsequently, decay to the LSP emitting a Higgs. This is not the usual assumption in experimental papers, where simplified models, with the gluino decaying 100\% to the LSP and jets, are considered, and therefore all bounds should be reinterpreted for this particular case. Moreover, a characteristic signature of this kind of spectra would be a gluino, with different decay patterns, having Higgses in the cascade.
 
Stops are the second possibility of colored particles that could be discovered at the LHC. We have presented a typical benchmark spectrum with stops masses of  $\sim$1.6 TeV in Tab.~\ref{tab:HMSSM-NUGM}. This ``light" stop evades the current bounds because the amount of missing energy is too small to trigger on the event~\cite{Aad:2020sgw}. In order to discover stops in the case where the splitting between the stop and the LSP is around 500 GeV, we would need different techniques for this compressed situation.
 Whereas in the benchmark shown in Table~\ref{tab:HMSSM}, the lightest stop mass is $\sim$ 2 TeV, well above the current LHC bound~\cite{Aad:2020sgw}.  This bound will be improved in the HL-LHC run, and hopefully it will reach a 2 TeV stop mass.

\section{Conclusion}
\label{sec:conclusion}
In this paper we focussed on the appealing possibility that a nearly pure Higgsino with a mass $\sim$1.1 TeV be the LSP, and therefore constitutes the DM of our universe. We have done so in the context of the MSSM, with supersymmetry breaking triggered by gravitational interactions at the (high) unification scale $M\simeq 2\times 10^{16}$ GeV, in which case the $\mu$-parameter of the superpotential can be generated by Higgs interactions in the K\"ahler potential, the so-called Giudice-Masiero mechanism. 

In particular we have considered two classes of models: i) Models with universal Higgs masses, i.e.~models where all scalar masses, for sfermions and Higgses, are equal at the unification scale, and ii) Models with non-universal Higgs masses, i.e.~models where the common mass in the Higgs sector is in general different from the common mass in the sfermion sector. In both classes of models we have considered the cases of universal gaugino masses (i.e.~all gaugino Majorana masses equal at the unification scale) and the cases where only the electroweakino masses are unified at the unification scale, while the gluino mass is put at its experimental lower bound. 

In view of the strong experimental bounds on the mass of supersymmetric particles we have focussed this work in the search of spectra which can be at reach of the future LHC runs, or future high energy colliders. Notice that, as the LSP is a Higgsino with a mass equal to 1.1 TeV, this means that all other supersymmetric particles are heavier and with no easy detection, in agreement with the recent experimental results on supersymmetric searches.

The conclusion for the neutralino/chargino sector is pretty general and model independent. The two lightest neutralinos and the lightest chargino are quasi degenerate at a mass $\sim$1.1 TeV, with a splitting of order a few GeV. There is more freedom on the masses for the other electroweakinos, but using the bound from XENON1T as a guidance, we have a heavy neutralino, with a mass larger than 1.5 TeV and the heaviest neutralino and chargino, almost degenerate, with a mass larger than 2.7 TeV.  

For the sfermion sector, models with universal Higgs mass have sfermions heavier than $\sim$10 TeV, depending on the gluino mass, and thus the only observable particle, in this class of models, can eventually be the gluino for models with non-universal gaugino masses. Models with non-universal Higgs mass can have squarks in the (few) TeV range, depending on the gluino mass. For heavy gluinos the lightest sfermion is the right-handed stau.
For light gluinos, the renormalization effects are milder, and there can exist TeV squarks in the third generation, and sizable mixing in the third generation squarks, such that the experimental value of the Higgs mass can be easily accomodated. In this case the lightest sfermion is the lightest stop.

In summary, the main purpose of this work was to corner spectra which could be at reach of future collider searches. In short, we have found that the lighter particles can be, apart from the Higgsino which is the LSP, the gluino, the right-handed stau and the lightest stop. The prospect for discovering the Higgsino are at a future 100 TeV collider and by direct search experiments at the XENON-nT or LZ experiments. The gluino can be discovered at the LHC, if it is light enough. The fact that there is a neutralino (mostly bino) with a mass between the gluino and the LSP translates into additional decay channels for the gluino which should be incorporated in the codes used by the experimental programs. Finally we have a model with the  lightest stop with a mass $\sim$1.3 TeV which could be discovered in the HL-LHC run.

\section*{Acknowledgments}
The work of AD is partly supported by the National Science Foundation under grant PHY-1820860. The work of MQ is partly supported by Spanish MINEICO under Grant FPA2017-88915-P, by the Catalan Government under Grant 2017SGR1069, and by Severo Ochoa Excellence Program of MINEICO under Grant SEV-2016-0588).


\end{document}